\documentclass[epj]{svjour}

\usepackage{graphics}
\usepackage{hyperref}
\usepackage{amsmath}
\usepackage{epsfig}

\newlength{\myVSpace}
\newcommand\xstrut{\raisebox{.5\myVSpace}{\rule{0pt}{\myVSpace}}}

\begin{document}

\title{Search for the $\Theta^+(1540)$ in lattice QCD}

\author{Christian Hagen, Dieter Hierl \and Andreas Sch\"afer}

\institute{Institut f\"ur theoretische Physik, Universit\"at Regensburg, D-93040 Regensburg, Germany}

\abstract{
We report on a study of the pentaquark $\Theta^+(1540)$, using a variety of different interpolating fields. We use Chirally Improved fermions in combination with Jacobi smeared quark sources to improve the signal and get reliable results even for small quark masses. The results of our quenched calculations, which have been done on a $12^3\times24$ lattice with a lattice spacing of $a=0.148~{\rm fm}$, do not provide any evidence for the existence of a $\Theta^+$ with positive parity. We do observe, however, a signal compatible with nucleon-kaon scattering state. For the negative parity the results are inconclusive, due to the potential mixture with nucleon-kaon and $N^*$-kaon scattering states.
}

\maketitle

\section{Introduction}
\label{intro}

The possible discovery of the $\Theta^+(1540)$ by the LEPS Collaboration at SPring-8 \cite{Nakano:2003qx} 
has initiated great interest in exotic baryons. Since then, there has been a large number of experiments 
that have confirmed this result \cite{Stepanyan:2003qr}
-\cite{Troyan:2004wp} but also about the same number that could not confirm it 
\cite{Bai:2004gk}-\cite{Armstrong:2004gp}. Presently the experimental situation is highly unsettled 
\cite{:2005gy,Ricciardi:2005sn}.\newline
Also lattice QCD has not been able to give a conclusive answer so far to the question whether there might exist a bound $\Theta^+$ state or not \cite{Csikor:2003ng}-\cite{Holland:2005yt}.\newline
To confirm or disprove the existence of the $\Theta^+$ by lattice calculations is a huge challenge. 
If a clear conclusion of the  could be reached and would eventually be experimentally confirmed, 
this would give a substantial boost to lattice QCD. Therefore, many groups have started to work on this problem. 
Here we present our first results using some interpolators which have not been tested on the lattice yet.\newline
In our calculations we perform a qualitative study using different types of spin-$\frac{1}{2}$ operators with 
the quantum numbers of the $\Theta^+$. We compute all cross correlators and use the variational method 
\cite{Michael:1985ne,Luscher:1990ck} to extract the lowest lying eigenvalues. These are used to create 
effective mass plots for a comparison to the $N$-$K$ scattering state which we computed separately on 
the same lattice.

\section{Details of the calculation}
\label{settings}

We considered the following interpolating fields as basis for our correlation matrix:
\begin{itemize}

\item
Currents suggested by Csikor/Fodor \cite{Csikor:2003ng}:

\begin{equation} \label{fodor12}
I_{0/1}=\epsilon_{abc}[u_a^T C \gamma_5 d_b] 
\{u_e \bar{s}_e i \gamma_5 d_c \mp (u \leftrightarrow d)\}
\end{equation}

\item
Currents suggested by Sasaki \cite{Sasaki:2003gi}:

\begin{eqnarray}
\label{sasaki1}
\Theta_+^1&=&\epsilon_{abc}\epsilon_{aef}\epsilon_{bgh}
(u_e^T C d_f)(u_g^T C \gamma_5 d_h )C\bar{s}_c^T
\\\nonumber\\
\label{sasaki2}
\Theta_{+,\mu}^2&=&\epsilon_{abc}\epsilon_{aef}\epsilon_{bgh}
(u_e^T C \gamma_5 d_f)(u_g^T C \gamma_5 \gamma_\mu d_h )C\bar{s}_c^T
\\\nonumber\\
\label{sasaki3}
\Theta_{-,\mu}^3&=&\epsilon_{abc}\epsilon_{aef}\epsilon_{bgh}
(u_e^T C d_f)(u_g^T C \gamma_5 \gamma_\mu d_h )C\bar{s}_c^T
\end{eqnarray}

\item
Currents which are suggestions by L. Ya. Glozman \cite{Glozman:2003sy}, however, using only s-wave quarks instead of a mixture of s-wave and p-wave quarks in (\ref{glozman2}) and (\ref{glozman3}):

\begin{eqnarray}
\label{glozman2}I_{\mu}&=&(\delta_{ae}\delta_{bg}+\delta_{be}\delta_{ag})\epsilon_{gcd}\\
&\times&\left(
\begin{array}{c}
u_a^T C u_b\\
\frac{1}{\sqrt{2}} (u_a^T C d_b + d_a^T C u_b )\\
d_a^T C d_b
\end{array}
\right)\nonumber\\
&\times&\left(
\begin{array}{c}
d_c^T (C \gamma_\mu ) d_d\\
\frac{1}{\sqrt{2}} (u_c^T (C \gamma_\mu ) d_d + d_c^T (C \gamma_\mu ) u_d)\\
u_c^T (C \gamma_\mu ) u_d
\end{array}
\right)
C\bar{s}_e^T\nonumber
\end{eqnarray}

\begin{eqnarray}
\label{glozman3}I&=&(\delta_{ae}\delta_{bg}+\delta_{be}\delta_{ag})\epsilon_{gcd}\\
&\times&\left(
\begin{array}{c}
u_a^T (C \gamma_\mu \gamma_5 ) u_b\\
\frac{1}{\sqrt{2}} (u_a^T (C \gamma_\mu \gamma_5 ) d_b + d_a^T (C \gamma_\mu \gamma_5 ) u_b )\\
d_a^T (C \gamma_\mu \gamma_5 ) d_b
\end{array}
\right)\nonumber\\
&\times&\left(
\begin{array}{c}
d_c^T (C \gamma_\mu ) d_d\\
\frac{1}{\sqrt{2}} (u_c^T (C \gamma_\mu ) d_d + d_c^T (C \gamma_\mu ) u_d)\\
u_c^T (C \gamma_\mu ) u_d
\end{array}
\right)
C\bar{s}_e^T\nonumber
\end{eqnarray}

\item
Other currents related to suggestions by L. Ya. Glozman \cite{Glozman:2003sy}, using only s-wave quarks and a factor $(\delta_{ae}\delta_{bg}-\delta_{be}\delta_{ag})$ instead the factor $(\delta_{ae}\delta_{bg}+\delta_{be}\delta_{ag})$ in \eqref{glozman1} and \eqref{glozman4}:

\begin{eqnarray}
\label{glozman1}I_{\mu}&=&(\delta_{ae}\delta_{bg}-\delta_{be}\delta_{ag})\epsilon_{gcd}\\
&\times&[u_a^T (C \gamma_\mu \gamma_5) d_b - d_a^T (C \gamma_\mu \gamma_5) u_b ]\nonumber\\
&\times&[u_c^T (C \gamma_5 ) d_d - d_c^T (C \gamma_5 ) u_d ]C\bar{s}_e^T\nonumber
\end{eqnarray}

\begin{eqnarray}
\label{glozman4}I_{\nu}&=&(\delta_{ae}\delta_{bg}-\delta_{be}\delta_{ag})\epsilon_{gcd}\\
&\times&\left(
\begin{array}{c}
u_a^T (C \sigma_{\mu \nu} ) u_b\\
\frac{1}{\sqrt{2}} (u_a^T (C \sigma_{\mu \nu} ) d_b + d_a^T (C \sigma_{\mu \nu} ) u_b )\\
d_a^T (C \sigma_{\mu \nu} ) d_b
\end{array}
\right)\nonumber\\
&\times&\left(
\begin{array}{c}
d_c^T (C \gamma_\mu ) d_d\\
\frac{1}{\sqrt{2}} (u_c^T (C \gamma_\mu ) d_d + d_c^T (C \gamma_\mu ) u_d)\\
u_c^T (C \gamma_\mu ) u_d
\end{array}
\right)
C\bar{s}_e^T\nonumber
\end{eqnarray}

\item
While the interpolating fields \eqref{fodor12} to \eqref{sasaki3} were already used in other lattice calculations the interpolators \eqref{glozman2} to \eqref{glozman4} have not been tested on the lattice yet.

\end{itemize}

In order to get only spin-$\frac{1}{2}$ pentaquarks, we have to project the spin. This is done using the spin projection operator for a Rarita-Schwinger field \cite{lurie1968}. The corresponding spin-$\frac{1}{2}$ state can be projected by applying the projection operator
\begin{equation}
P_{\mu\nu}^{1/2} = g_{\mu\nu} - P_{\mu\nu}^{3/2}\;,
\end{equation}
where
\begin{equation}
P^{3/2}_{\mu \nu}(p)= g_{\mu \nu}-\frac{1}{3}\gamma_\mu\gamma_\nu-\frac{1}{3p^2}\left(\gamma\cdot p\,\gamma_\mu p_\nu+p_\mu\gamma_\nu\,\gamma\cdot p\right)\;.
\end{equation}
One can show that choosing the temporal components of the Lorentz indices at zero momentum is sufficient to get only spin-$\frac{1}{2}$ contributions.

Our baryon correlators are also projected to definite parity using the projection operator $P^\pm=\frac{1}{2}(1\pm\gamma_4)$. We obtain two matrices of correlators:
\begin{eqnarray}
C^+_{ij}(t) & = &  Z_{ij}^+ e^{-tE^+} + Z_{ij}^- e^{-(T-t)E^-} ,
\end{eqnarray}
when we project with $P^+$ and
\begin{eqnarray}
C^-_{ij}(t) & = &  - Z_{ij}^- e^{-tE^-} - Z_{ij}^+ e^{-(T-t)E^+} ,
\end{eqnarray}
when we use $P^-$ as shown in \cite{Sasaki:2001nf}. These two matrices are combined to
\begin{eqnarray}
C(t) & = & \frac{1}{2} \left( C^+(t) - C^-(T-t) \right) \; ,
\end{eqnarray}
to improve the statistics. This in combination with our quarks having antiperiodic boundary conditions in temporal direction allows us to separate the parity channels. The parity channel of the $\Theta^+$ is not known. There are conflicting theoretical predictions and no conclusive experimental data, so we have to look at both channels.\newline
For the interpolators (\ref{glozman2}) to (\ref{glozman4}) p-wave quarks\footnote{Quarks with an orbital p-wave excitation in their wave functions.} would be required. Since we do not have p-wave sources we had to adjust the color structure in the interpolators (\ref{glozman1}) and (\ref{glozman4}) to obtain a signal at all\footnote{We changed the color factor $(\delta_{ae}\delta_{bg}+\delta_{be}\delta_{ag})$ into $(\delta_{ae}\delta_{bg}-\delta_{be}\delta_{ag})$.}. By doing so one immediately sees that the interpolator (\ref{glozman1}) becomes, up to a constant factor, the same as the operator (\ref{sasaki2}). Performing a small test simulation on a few configurations, we find that the interpolator (\ref{glozman3}) is numerically the same as interpolator (\ref{sasaki1}). Therefore we exclude the interpolators (\ref{sasaki1}) and (\ref{sasaki2}) from our analysis.\newline
The interpolator \eqref{glozman1} contains two diquarks with $I=0$ and so it has isospin $I=0$. In contrast the interpolators \eqref{glozman2}, \eqref{glozman3} and \eqref{glozman4} are linear combinations of two diquarks with $I=1$. Thus they are a mixture of isospin $I=0$ and $I=2$ states.\newline
We also have to exclude the interpolators \eqref{fodor12} because the cross correlations between them and the other interpolating fields are more complex and thus more demanding in computation.\newline
Then we use the remaining five interpolators to calculate a cross correlation matrix $C_{ij}(t)$ which is then inserted into the generalized eigenvalue problem
\begin{eqnarray}
C_{ij}(t) v_j^{(k)}&=&\lambda^{(k)}(t) C_{ij}(t_0) v_j^{(k)}.
\end{eqnarray}
The solutions of this equation behave like
\begin{eqnarray}
\lambda^{(k)}(t)&\propto&\exp\left(-m^{(k)}(t-t_0)\right).
\end{eqnarray}
These eigenvalues are used to compute effective masses according to
\begin{equation}\label{effmass}
m_{eff}(t)=\ln\left(\frac{\lambda(t)}{\lambda(t+1)}\right).
\end{equation}
Ordering the five eigenvalues according to their absolute value the largest eigenvalue in the positive parity channel should give the $\Theta^+$ mass if $\Theta^+$ is a positive parity particle. The second largest eigenvalue in the negative parity channel should give the $\Theta^+$ mass, where the largest eigenvalue corresponds to the $N$-$K$ scattering state at rest.\newline
In our quenched calculation we use the Chirally Improved Dirac operator \cite{Gattringer:2000js,Gattringer:2000qu}. It is an approximate solution of the Ginsparg-Wilson equation \cite{Ginsparg:1981bj}, with good chiral behavior \cite{Gattringer:2003qx,Gattringer:2002xg,Gattringer:2002sb}. The gauge fields are generated with the L\"uscher-Weisz gauge action \cite{Luscher:1984xn,Curci:1983an} at $\beta = 7.90$. The corresponding value of the lattice spacing is $a=0.148~{\rm fm}$ as determined from the Sommer parameter in \cite{Gattringer:2001jf}. For the quark sources we use a Gaussian-type distribution generated with Jacobi smearing \cite{Gusken:1989ad,Best:1997qp}. The error bars are computed using the jackknife method. The s-quark mass is determined from the Kaon mass. The parameters of our calculation are collected in Table \ref{latticetable}.

\begin{table}
\begin{center}
\begin{tabular}{|c|c|}
\hline
size $L^3\times T$&$12^3\times24$\xstrut\\\hline
$\beta$&7.90\\\hline
$a~[{\rm fm}]$&0.148\\\hline
$a^{-1}~[{\rm MeV}]$\xstrut&1333\\\hline
$L~[{\rm fm}]$&$\approx1.8$\\\hline
$\#$conf N&100\\\hline
&0.04, 0.05, 0.06, 0.08,\\
quark masses $am_q$& 0.10, 0.12, 0.16, 0.20\\\hline
smearing parameters:&$n=18$, $\kappa=0.210$\\\hline
s-quark mass $am_s$&0.0888(17)\\\hline
\end{tabular}
\end{center}
\caption{Parameters of our calculations.}
\label{latticetable}
\end{table}

\section{Results}
\label{results}

\begin{figure*}
\begin{center}
\resizebox{0.85\textwidth}{!}{\includegraphics[clip]{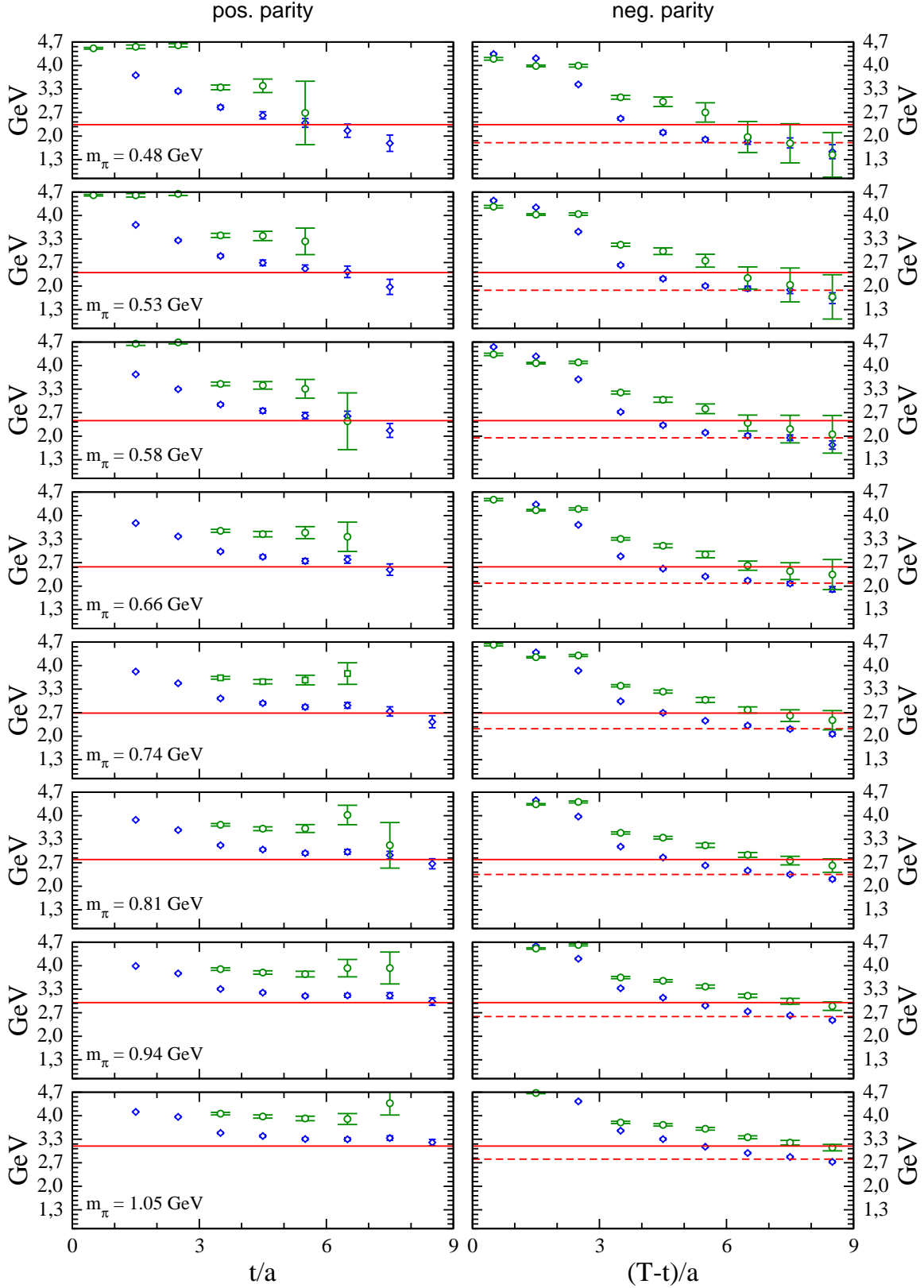}}
\end{center}
\caption
{Results from cross-correlation of our final set of interpolating fields, i.e., \eqref{sasaki3}, \eqref{glozman2}, \eqref{glozman3}, \eqref{glozman1} and \eqref{glozman4}. We show effective masses for the two lowest-lying eigenvalues computed according to Eq. \eqref{effmass}. The dashed line represents $M_N + M_K$ obtained from a separate calculation on the same lattice. The solid line is the energy for the smallest momentum calculated according to Eq. \eqref{relep}. For all our quarks we use Jacobi smeared Gaussian sources.}
\label{fig:1}       
\end{figure*}

\begin{figure}
\begin{center}
\resizebox{0.49\textwidth}{!}{\includegraphics[clip]{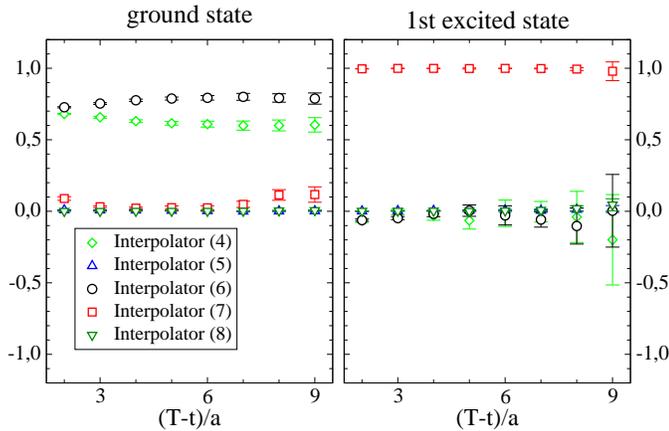}}
\end{center}
\caption
{Plot of the eigenvector components for $am_q=0.08$ which corresponds to the pion mass $m_\pi = 0.66$GeV. 
The left hand side plot shows the eigenvector components of the negative parity ground state, the right 
hand side plot is for the negative parity 1st excited state. The eigenvectors are obviously stable and  
different. Thus, we conclude that we observe to independent states.}
\label{fig:2}       
\end{figure}

The results of our calculations are shown in Fig. \ref{fig:1}, where we plot the effective masses of the two lowest lying states of both parity channels obtained with the cross-correlation technique. These states are approaching a possible plateau very slowly as we expected, since states consisting of five quarks are very complicated and therefore should contain a large number of excited states which have to die out before the effective mass reaches a plateau. We use in addition to the cross-correlation technique Jacobi smeared Gaussian quark sources for all our quarks to improve the signal for the lowest lying states.\newline
The lower dashed horizontal line in the negative parity channel is the sum of the nucleon and kaon mass at rest in the ground state obtained from a separate calculation on the same lattice. Since we project the final state to zero momentum a scattering state can also be a two particle state where the two particles have the same but antiparallel momentum, i.e. $\vec{p_N}=-\vec{p_K}$. We use the relativistic $E$-$p$-relation to calculate the energy of such states,
\begin{eqnarray}
\label{relep}
E&=&\sqrt{p^2+m_N^2}+\sqrt{p^2+m_K^2},
\end{eqnarray}
where the smallest momentum is $2\pi/L\approx700~{\rm MeV}$ on our lattice. In Fig. \ref{fig:1} this energy is represented by the upper horizontal line.\newline
In contrast to similar calculations we performed for ordinary mesons and baryons we do not find clean mass plateaus in 
this case. This implies also that we cannot perform any systematic chiral extrapolation. The signals we get are 
compatible with $N$-$K$ continuum states, but do not unambiguously identify them. For the latter we would have 
needed results for different volumes, which would have been very expensive. Furthermore, we are not interested in identifying such a continuum state, but only want to know whether there is any indication 
for a novel bound state.\newline
In the negative parity channel we find effective mass pla-teaus which are consistent with the lowest $N$-$K$ scattering states. We find that the second state is noisy but within errors consistent with the energy in Eq. \eqref{relep}. Therefore, it is most likely that we do not observe a $\Theta^+$ state in the negative parity channel. Naturally, if the $\Theta^+$ were broad, which implied that it  would mix strongly with the continuum our conclusion would be weakened. However, for the $\Theta^+$ this possibility is excluded experimentally.\newline
From the composition of the eigenvectors, shown in Fig. \ref{fig:2}, we conclude that the two lowest 
lying states are really independent of each other. One can also see that some of the interpolators which have 
not been used in previous studies actually give large contributions to the low lying states.\newline
In the positive parity channel one expects to find either a bound $\Theta^+$ or an excited $N$-$K$ scattering 
state. For such an excited state there are several possibilities, e.g., $N^*$-$K$, or $N$-$K$ with relative 
angular momentum, and so on.\newline
On the positive parity side, we also show the two lowest lying states obtained from our calculation. 
Both of them are too heavy to have anything to do with the $\Theta^+$. They probably correspond to excited 
$N$-$K$ scattering states. If there were a signal belonging to the $\Theta^+$ it is supposed to lie below 
the solid line assuming that the chiral extrapolation of the $\Theta^+$ does not lead to dramatic effects 
below our smallest quark mass. (A serious chiral extrapolation would require cleaner mass plateaus as noted 
above.)

\section{Conclusion}
\label{conclusion}

In this article we present the results of a pilot study for $\Theta^+$ searches on the lattice using different types of operators. We find that for negative parity our results are in good agreement with a $N$-$K$ scattering state in the ground state and a quite noisy signal for the first excited state.\newline
For positive parity we find states which are typically more than $500~{\rm MeV}$ heavier than the $\Theta^+$ 
and thus not compatible with a $\Theta^+$ mass of $1540~{\rm MeV}$. \footnote{We assume that the $\Theta^+$ extrapolates smoothly in both, the chiral and the continuum limit.}\newline
In our calculation we only used one lattice volume. If we had found a $\Theta^+$ candidate, for the final distinction between a bound or a scattering state we would have had to study the volume dependence. As it stands, there is no motivation to do so.\newline 
Up to now our calculation do not show any hints for a $\Theta^+$ in the quenched approximation with chiral fermions using in addition to standard ones the new interpolators \eqref{glozman2}-\eqref{glozman4}.

\section*{Acknowledgments}
\label{ack}
This work was funded by DFG and BMBF. We thank Ch. Gattringer, T. Burch and L.~Ya. Glozman for very helpful discussions and their support. All computations were done on the Hitachi SR8000 at the Leibniz-Rechenzentrum in Munich, on the JUMP cluster at NIC in J\"ulich and at the Rechenzentrum in Regensburg.

\end{document}